\documentclass[twocolumn]{aastex631}
\received{x}
\revised{x}
\accepted{x}
\submitjournal{ApJ}
\usepackage{fancyvrb} 
\usepackage{breakurl}
\usepackage{ulem}
\VerbatimFootnotes    

\begin{document}

\newcommand{\vdag}{(v)^\dagger}
\newcommand\aastex{AAS\TeX}
\newcommand\latex{La\TeX}

\shorttitle{nu sources}
\shortauthors{Kun et al.}
\graphicspath{{./}{figures/}}

\title{Multiwavelength search for the origin of IceCube's neutrinos}

\correspondingauthor{Emma Kun}
\email{kun.emma@csfk.org}

\author[00000-0003-2769-3591]{Emma Kun}
\affiliation{Konkoly Observatory, Research Centre for Astronomy and Earth Sciences,  MTA Centre of Excellence, Budapest, Konkoly Thege Miklós út 15-17., H-1121, Hungary}
\affiliation{CSFK, MTA Centre of Excellence, Budapest, Konkoly Thege Miklós út 15-17., H-1121, Hungary}

\author[0000-0001-5607-3637]{Imre Bartos}
\affiliation{Department of Physics, University of Florida, PO Box 118440, Gainesville, FL 32611-8440, USA}

\author[0000-0002-1748-7367]{Julia Becker Tjus}
\affiliation{Theoretical Physics IV: Plasma-Astroparticle Physics, Faculty for Physics \& Astronomy, Ruhr University Bochum, 44780 Bochum, Germany}
\affiliation{Ruhr Astroparticle And Plasma Physics Center (RAPP Center), Ruhr University Bochum, 44780 Bochum, Germany}

\author[00000-0000-0000-0000]{Peter L.\ Biermann}
\affiliation{MPI for Radioastronomy, 53121 Bonn, Germany}
\affiliation{Department of Physics \& Astronomy, University of Alabama, Tuscaloosa, AL 35487, USA}

\author[0000-0002-5605-2219]{Anna Franckowiak}
\affiliation{Faculty of Physics and Astronomy, Astronomical Institute (AIRUB), Ruhr University Bochum, 44780 Bochum, Germany}

\author[0000-0001-6224-2417]{Francis Halzen}
\affiliation{Dept. of Physics, University of Wisconsin, Madison, WI 53706, USA}

\begin{abstract}
The origin of astrophysical high-energy neutrinos detected by the IceCube Neutrino Observatory remains a mystery to be solved. In this paper we search for neutrino source candidates within the $90$\% containment area of $70$ track-type neutrino events recorded by the IceCube Neutrino Observatory. By employing the \textit{Fermi}-LAT 4FGL-DR2, the \textit{Swift}-XRT 2SXPS and the CRATES catalogs, we identify possible gamma, X-ray and flat-spectrum radio candidate sources of track-type neutrinos. We find that based on the brightness of sources and their spatial correlation with the track-type IceCube neutrinos, the constructed neutrino samples represent special populations of sources taken from the full \textit{Fermi}-LAT 4FGL-DR2/\textit{Swift}-XRT 2SXPS/CRATES catalogs with similar significance ($2.1\sigma$, $1.2\sigma$, $2\sigma$ at $4.8~\mathrm{GHz}$, $2.1\sigma$ at $8.4~\mathrm{GHz}$, respectively, assuming 50\% astrophysical signalness). After collecting redshifts and deriving sub-samples of the CRATES catalog complete in the redshift--luminosity plane, we find that the 4.8 GHz ($8.4$~GHz) sub-sample can explain between 4\% and 53\% ($3$\% and $42$\%) of the neutrinos (90\% C.L.), when the probability to detect a neutrino is proportional to the ($k$-corrected) radio flux. The overfluctuations indicate that a part of the sample is likely to contribute and that more sophisticated schemes in the source catalog selection are necessary to identify the neutrino sources at the $5\sigma$ level. Our selection serves as a starting point to further select the correct sources.
\end{abstract}

\section{Introduction}
\label{sec:intro}

High-energy cosmic rays (HECRs) are energetic subatomic particles with a non-thermal energy spectrum that ranges from GeV up to at most ZeV energies \citep[e.g.,][]{Auger2010,telarray2012,Auger2017,Auger2019,Biermann2016,Biermann2018}. The detection of high-energy neutrinos is an unambiguous way of pinpointing the high-energy particle accelerators, a necessary precursor to ultra-high-energy cosmic ray (UHECR) particles in the sky \citep{Gaisser1990}, that cannot be revealed in cosmic rays alone. The reason is that HECRs scatter in Galactic and intergalactic magnetic fields \citep{Dermer2009,Tjus2020}, losing directionality. An alternative is the detection of high-energy gamma-rays arising from hadronic interactions. These are, however, absorbed in interactions with the CMB photons, so the high-energy gamma-ray horizon at TeV energies is constrained to the universe at $z<1$, and at PeV energies it is constrained to our Galaxy. In addition, the leptonic production of high-energy gamma-rays makes an interpretation of the photon spectrum alone much more difficult.

Neutrinos are produced in high-energy \mbox{(photo-)hadronic} interaction processes via the production of pions and kaons, which subsequently decay to produce high-energy neutrinos and muons, where the latter further contribute to the neutrino flux \citep{Gaisser1995,Becker2008}. High-energy gamma-rays are always co-produced via the decay of neutral pions. The IceCube Neutrino Observatory detects cosmic high-energy neutrinos with energies between $\sim 100$~GeV and $\sim 10$~PeV. Among millions of atmospheric neutrinos and hundreds of billions of cosmic-ray muons, about $200$/year cosmic neutrinos for an $E^{-2.5}$ spectrum \citep[][]{Aartsen2015} are detected via data reduction by introducing intelligent physics-based cuts to remove a large portion of the atmospheric particles \citep{Aartsen2013Science,Aartsen2013PRL,Icecube2014,Icecube2016,Icecube2017,Aartsen2020}.

After their first detection by the IceCube Neutrino Detector \citep{Aartsen2013Science}, high-energy astrophysical neutrinos quickly became powerful probes of astrophysics \citep[e.g.][]{Ahlers2018}. In 2018, multi-messenger observations of the blazar TXS 0506+056 were reported \citep{ICTXS2018a}. The high-energy neutrino event IceCube-170922A detected at 22/9/2017 was in temporal and positional coincidence with a strong gamma-ray flare shown by TXS 0506+056 detected by \textit{Fermi}-LAT at GeV energies. Knowing where to look, the IceCube Collaboration analyzed archival track-type data and found $3.5\sigma$ evidence for a neutrino flare in the direction of TXS 0506+056 \citep{ICTXS2018b} during a low gamma state of TXS 0506+06 \citep{Garrappa2019}, that lasted about half a year and was of significantly lower energy as compared to the IceCube event IceCube-170922A. Since 2019 June, neutrino events with high chance to be astrophysical and detected in the framework of the IceCube Realtime Alert System \citep{Aartsen2017realtime} are distributed to the Gold (signalness is above 50\%) or Bronze channel (signalness is below 50\% but above 30\%). The signalness is the probability that an observed neutrino is of astrophysical ($s =
1$) or atmospheric ($s = 0$) origin \citep[see its definition in][]{Aartsen2017realtime}.

In general, radio-loud active galactic nuclei (AGN) are among the primary candidate sources of IceCube neutrinos. A prime candidate is their blazar sub-class. These sources usually show low-energy and high-energy bumps coexisting in their spectral energy distribution (SED). Throughout the paper we refer to blazars as the subclass of AGN that point their jet close to our line-of-sight, according to the unification scheme of radio-loud AGN \citep[][]{Urry1995}. The flat spectrum at the $5$~GHz range of radio frequencies is indicative of these sources pointing to us and boosting the emission \citep{Biermann1981,Gregorini1984}.

High-energy neutrinos can be produced in the interaction of cosmic rays with either a photon or a gas target. Here, threshold of the kinematic energy of a cosmic-ray proton for interactions with the gas is at $E_p\gtrsim 280$~MeV. Interactions with a photon target, on the other hand, is typically constrained to higher energies, as its threshold is dominated by the production of the delta-resonance at $E\gtrsim 1.3$~GeV. As photon targets are typically most dense for thermal emission spectra in the infrared to optical energy range, these interactions might be very effective in certain regions like close to the accretion disk of the supermassive black hole, but the energy spectrum will be limited to energies above $\sim 10^{18}$~eV for the Bethe–Heitler pair production and above $\sim 5\times10^{19}$~eV for the photopion production. 

Two photohadronic processes are of astrophysical interest as mentioned, the Bethe–Heitler pair production and the photopion production, such that neutrinos are produced only by the latter process. Enhanced gamma-ray emission from the decay of neutral pions is bound to accompany the emission of neutrinos born in charged current pion production. However, no statistically significant association of neutrino events with an ensemble gamma-ray loud blazars has been found yet \citep[e.g.][]{Aartsen2017,Neronov2017,OSullivan2019,Aartsen2020}. Moreover, at lower energies there is more flux in the IceCube diffuse neutrino component than that could be expected from the \textit{Fermi} diffuse gamma-ray sky if they have the same sources \citep{7d5yearsicecubeshowers}. 

Independently, it has been shown that \textit{Fermi}-LAT resolved blazars contribute only up to $\sim5-30\%$ of the IceCube cumulative neutrino flux \citep[][]{Aartsen2017,Pinat2017,Hooper2019}. Recently \citet{Bartos2021} found blazars (mainly FSRQs, based on their assumption of the blazar density) cannot contribute more than $11$\% to the total diffuse neutrino flux (90\% confidence level, C.L.). We note these results are model dependent as regards the diffuse backgrounds.

Hadronic gamma-rays produced in photon-rich environments, such as AGN, may lose energy through electromagnetic cascading, which results in a high flux emitted in the hard X-ray to MeV range \citep[e.g.][]{Petropoulou2015a,Petropoulou2015b,Murase2018,Rodrigues2019,Halzen2020}. \citet{Petropoulou2015b} had shown that if relativistic protons interact with the synchrotron blazar photons producing gamma-rays through photopion processes, despite being a subdominant proton cooling process (for protons satisfying the threshold condition for photopion production), the typical blazar SED should have an emission feature due to the synchrotron emission of Bethe-Heitler pairs in the energy range $40$~keV--$40$~MeV. Therefore, typical neutrino emitters would show this third, although less energetic bump between the typical low-energy and high-energy bumps of blazar SEDs \citep[e.g.][]{Abdo2010}. 

It is a nontrivial task to find a well-established connection between the radio flux and neutrino flux in AGN, mainly because neutrino production zones may be realized in different parts of an AGN (central or extended parts of the galaxy), and the resolution of the instruments, especially at high energies, is somewhat poor. While a significant connection is not widely concluded yet, catalog searches and studies of particular individual radio-loud sources associated with neutrinos \citep[e.g.][]{Kadler2016,Kun2019,Britzen2019,Ros2020,Britzen2021,Kun2021,Rodrigues2021} deliver more and more results that suggest radio observations play an important role in the understanding of the sources of astrophysical neutrinos.

One way to study the connection between radio observations of AGN and neutrino data observed with IceCube is the search for correlations between neutrinos that have a high probability of being astrophysical and individual sources or source catalogs. By analyzing public muon-track data from the $7$ years of IceCube observations against the $8$~GHz VLBI data of $3,411$ radio-loud AGN in the Radio Fundamental Catalog, \citet{Plavin2021} found a connection between the neutrinos and the radio sources at a $4.1\sigma$ significance level, combining the post-trial significance of their study and the results in \citet{Plavin2020}.
\citet{Hovatta2021} studied association of IceCube neutrinos with radio sources observed at Owens Valley (at $15$~GHz) and Metsähovi Radio Observatories (at $36.8$~GHz) by identifying sources in their radio monitoring sample that are positionally consistent with IceCube high-energy neutrino events. According to their results, observations of large-amplitude radio flares in a blazar at the same time as a neutrino event are unlikely to be a random coincidence at $2\sigma$ level. 

Another way to study the connection between radio observations of AGN and neutrino data taken with IceCube is the concept of stacking, see \citet{AMANDA2006} for the introduction of the method of source stacking, for the first time in neutrino astronomy. \citet{Zhou2021} studied the connection between $10$ year of IceCube muon-track data (with $1,134,450$ tracks, taken between April 2008 and July 2018), and $3,388$ radio-bright AGN at $8$~GHz selected from the Radio Fundamental Catalog. Using the unbinned maximum likelihood method they analysed the neutrino data in the position of radio-bright AGN. Their results imply the $3,388$ radio-bright AGN can account for at most $30$\% ($95$\% C.L.) of the flux of neutrino tracks in the $10$ years catalog of IceCube. \citet{Desai2021} presented a correlation analysis between $15$~GHz radio observations of AGN reported in the MOJAVE XV. Catalog and $10$ years of IceCube detector data. They did not find a significant correlation between them.

In this paper we carry out a multiwavelength search of sources within the $90$\% containment area of $70$ IceCube track-type neutrino events. After the Introduction in Section \ref{sec:intro}., in Section \ref{sec:matches}. we design an analysis to match the $70$ track-type neutrino events published by IceCube and the \textit{Fermi}-LAT 4FGL-DR2 catalog, the \textit{Swift}-XRT 2SXPS catalog and the CRATES Catalog. We perform a randomness test of the average brightness of the candidate samples in Section \ref{sec:randomness}. We investigate neutrino-radio statistical correlations in Section \ref{sec:nuradio}. by deriving number counts of the CRATES catalog after collecting redshifts of more then $3,000$ radio sources, and by defining a 4.8 GHz and a 8.4 GHz sub-sample that are complete in radio luminosities. Then we estimate how many neutrinos can be explained by the above sub-samples, assuming the probability to detect a neutrino is assumed to be proportional to the radio flux. We summarize our results in Section \ref{sec:results}. and discuss them in Section \ref{sec:discussion}. We give our
conclusions in Section \ref{sec:conclusion}.

\section{Point sources within 90\% containment of IceCube neutrinos}
\label{sec:matches}

\begin{figure*}
    \centering
    \includegraphics[width=0.85\textwidth]{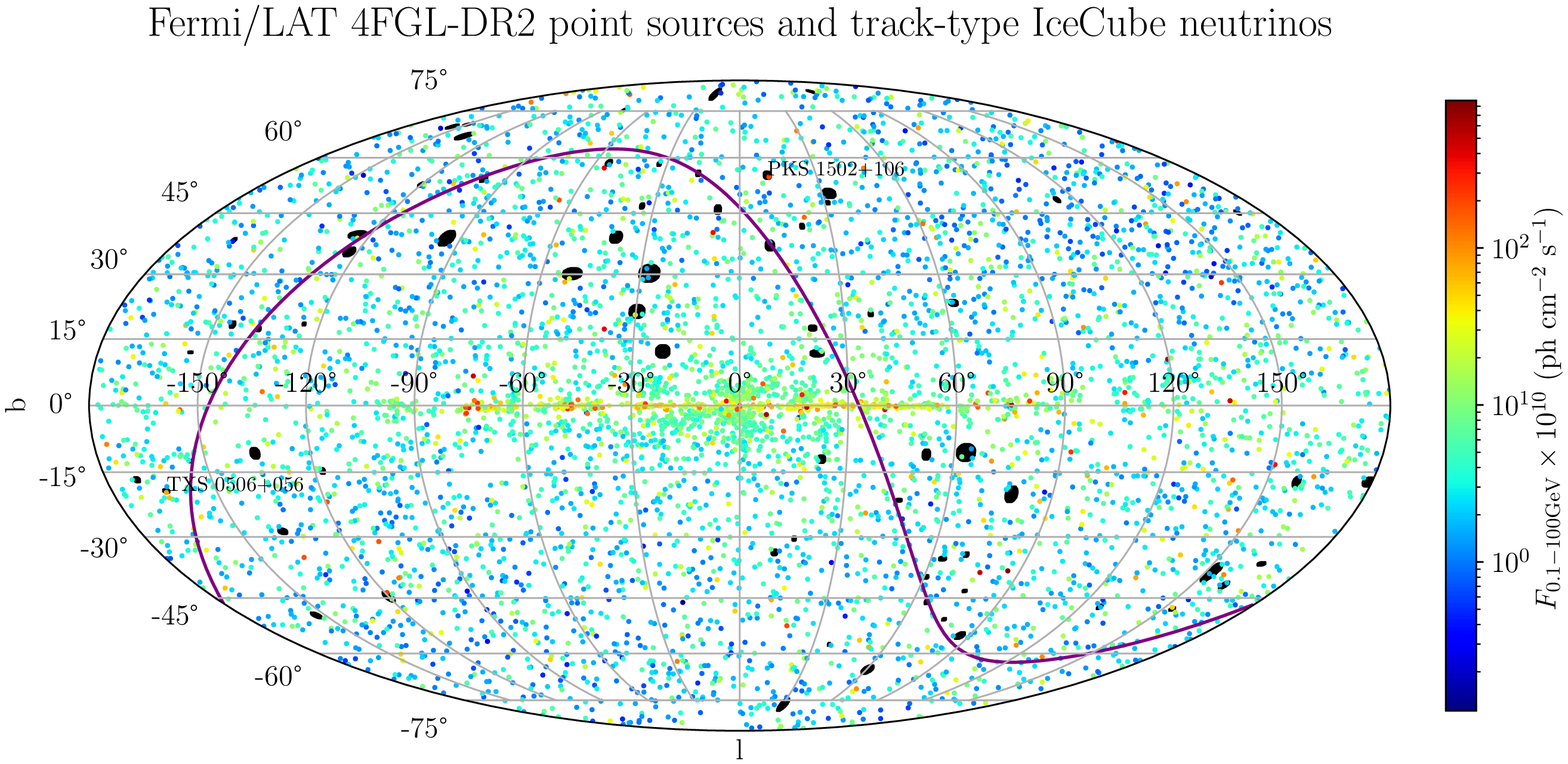}
    \caption{Galactic sky-position ($l$,$b$) and $0.1$ GeV-$100$ GeV flux ($F_{0.1-100 \mathrm{GeV}}$) of the \textit{Fermi}-LAT 4FGL-DR2 point sources (colored dots), overlaid on the list of the 70 track-type neutrino detection by the IceCube Neutrino Detector (black filled ellipses) compiled by \citet{Giommi2020}. The map is shown in Mollweide projection. The equator is plotted by a purple continuous line. TXS 0506+056 and PKS 1502+106 are marked on the map. The $7$ brightest sources (with $F_{0.1-100 ~\mathrm{GeV}} > 10^{-7}~ph~cm^{-2}s^{-1}$) are not shown to enhance the visibility of the flux-diversity of the sources.}
    \label{fig:gamma}
\end{figure*}

\begin{deluxetable*}{ccccccccc}
\tablenum{1}
\tablecaption{Excerpt of the neutrino source candidates from $10$-year \textit{Fermi}-LAT Point Source Catalogue (4FGL-DR2) ($n=29$, out of $5,787$). Neutrino ID (1), detection time (2), arrival direction of neutrino (90\% containment, 3-4), 4FGL-DR2 source ID (5), equatorial coordinates (J2000) (6-7), predicted number of photons in the $50$~MeV--$1$~TeV energy range (8), flux between $0.1$--$100$~GeV (9). The full table is available in electronic format \href{https://konkoly.hu/staff/kun.emma/table1_4fgl_fulltable.txt}{here}.\label{table:fermi}}
\tablewidth{0pt}
\tablehead{
\colhead{$ID_{\nu}$}  & \colhead{$t_{\nu}$} & \colhead{$RA_{\nu}$} & \colhead{$DEC_{\nu}$} & \colhead{$ID_\mathrm{4FGL}$} & \colhead{$RA_\mathrm{4FGL}$} & \colhead{$DEC_\mathrm{4FGL}$} & \colhead{$N_{pred}$} & \colhead{$F_{0.1-100 GeV}$}\\
 & (MJD) & ($^\circ$) & ($^\circ$) & & ($^\circ$) &  ($^\circ$) & &  ($~ph~cm^{-2}s^{-1}$)}
 \decimalcolnumbers
\startdata
IC-100710A & 55387.54 & $307.53^{+2.70}_{-2.28}$ & $21.00^{2.25}_{-1.56}$ & 4FGL J2030.5+2235 & 307.63 & 22.59 & 117.0 & 1.36e-10\\
IC-100710A & 55387.54 & $307.53^{+2.70}_{-2.28}$ & $21.00^{2.25}_{-1.56}$ & 4FGL J2030.9+1935 & 307.74 & 19.60 & 526.8 & 8.24e-10\\
IC-110610A & 55722.43 & $272.49^{+1.23}_{-1.19}$ & $35.55^{0.69}_{-0.69}$ & 4FGL J1808.2+3500 & 272.07 & 35.01 & 730.8 & 3.72e-10\\
IC-110610A & 55722.43 & $272.49^{+1.23}_{-1.19}$ & $35.55^{0.69}_{-0.69}$ & 4FGL J1808.8+3522 & 272.22 & 35.38 & 305.6 & 1.91e-10\\
IC-110930A & 55834.45 & $267.23^{+2.09}_{-1.55}$ & $-4.41^{0.59}_{-0.86}$ & 4FGL J1744.2-0353 & 266.05 & -3.89 & 740.6 & 3.90e-10\\
\enddata
\end{deluxetable*}

The neutrino list compiled by \citet{Giommi2020} summarizes 70 public IceCube high-energy cosmic neutrino tracks (recorded between 2009-08-13 and 2019-07-30) that are well reconstructed and off the Galactic plane. We use this sample to find neutrino candidate sources. Given that the errors on the neutrino sky positions are asymmetric, the center of their error-ellipses were shifted in the direction of the most significant errors by an amount equal to half the difference between the larger and smaller errors, and by setting the major and minor axes equal to the sum of the two asymmetrical errors.

\subsection{Gamma-ray neutrino source candidates in the \textit{Fermi}-LAT 4FGL-DR2 catalog}
\label{fermisearch}

The \textit{Fermi Gamma-ray Space Telescope} is one of the most successful telescopes in the gamma range. Its main instrument, the Large Area Telescope (LAT) is sensitive between $20$~MeV and $300$~GeV energies, though measurements are made up to $1$~TeV. The 4FGL-DR2 catalog includes $5,787$ gamma sources based on $10$ years of survey data in the $50$~MeV--$1$~TeV energy range \citep{Fermi4FGLDR2_2020}, being an incremental version to the $8$~years 4FGL catalog \citep{Fermi4FGL2020}. More than 3,413 of the identified or associated sources are blazars in the \textit{Fermi}-LAT 4FGL-DR2 catalog \citep{Fermi4FGL2020,Fermi4FGLDR2_2020}. We plot this catalog in Fig. \ref{fig:gamma} overlaid on the $90$\% containment area of the IceCube track-type neutrino events \citep{Giommi2020}. We found $29$ gamma-ray sources from the \textit{Fermi}-LAT 4FGL-DR2 catalog within the $90$\% containment area of track-type neutrinos, as summarized in Table \ref{table:fermi}, and from now on we refer to them as the gamma "$\nu$-sample". 
In Table \ref{table:results} we summarize the average, standard deviation, median, minimum and maximum values of the gamma flux in the $0.1$--$100$~GeV energy range of the sources in the full \textit{Fermi}-LAT 4FGL-DR2 catalog, and separately such values for the gamma $\nu$-sample. The average flux of the \textit{Fermi}-LAT $\nu$-sample emerged as $2.4\times 10^{-9} \mathrm{ph~cm}^{-2}s^{-1}$.

\begin{deluxetable}{ccccccc}
\tablenum{2}
\tablecaption{The number of sources (N), the average ($\bar{}$), standard deviation ($\sigma$), median (med), minimum (min) and maximum (max) values of the flux the sources in the total (T) \textit{Fermi}-LAT 4FGL-DR2 ($F_{0.1-100 ~\mathrm{GeV}}$), of the mean rate of sources in the full \textit{Swift}-XRT 2SXPS catalog ($R_m$), of the flux density at $4.8$ GHz ($S_{4.8}$) and $8.4$ GHz ($S_{8.4}$) of sources in the full CRATES catalog, as well as the same properties of the sub-samples of neutrino candidate sources (S) derived from the above catalogs by the list of IceCube track-type neutrino events compiled by \citet{Giommi2020}. \label{table:results}}
\tablewidth{0pt}
\tablehead{\multicolumn{7}{c}{\textit{Fermi}-LAT 4FGL -- $F_{0.1-100 ~\mathrm{GeV}}$ (in $ph~cm^{-2}s^{-1}$)}
}
\startdata
 & $N$ & $\bar{F}$ & $\sigma_F$ & med($F$) & min($F$) & max($F$)\\
& - & ($10^{-9}$) & ($10^{-9}$) & ($10^{-10}$) & ($10^{-11}$) & ($10^{-8}$)\\
\hline
T & $5,787$ & $1.5$ & $20.8$ & $3.2$ & $1.1$ & $135.2$\\
S & $29$ & $2.4$ & $5.6$ & $3.3$  & $3.9$ & $2.4$\\
\hline
\hline
\multicolumn{7}{c}{\textit{Swift}-XRT 2SXPS -- $R_m$ (in cts s$^{-1}$), $SNR\geqq10$}\\
 & $N$ & $\bar{R}_m$ & $\sigma_R$ & med($R_m$) & min($R_m$) & max($R_m$)\\
\hline
T & $9,097$ & $0.039$ & $0.077$ & $0.015$ & $0.002$ & $1.755$\\
S & $61$ & $0.049$ & $0.068$ & $0.027$ & $0.003$ & $0.3603$\\
\hline
\hline
\multicolumn{7}{c}{CRATES -- $S_{4.8}$ (in mJy)}\\
 & $N$ & $\bar{S}_{4.8}$ & $\sigma_{S4.8}$ & med(${S}_{4.8}$) & min(${S}_{4.8}$) & max(${S}_{4.8}$)\\
\hline
T & $11,131$ & $251$ & $905$ & $120$ & $65$ & $46,894$\\
S & $87$ & $544$ & $573$ & $232$ & $66$ & $2,530$\\
\hline
\hline
\multicolumn{7}{c}{CRATES -- $S_{8.4}$ (in mJy)}\\
 & $N$ & $\bar{S}_{8.4}$ & $\sigma_{S8.4}$ & med(${S}_{8.4}$) & min(${S}_{8.4}$) & max(${S}_{8.4}$)\\
\hline
T & $14,467$ & $149$ & $596$ & $65$ & $0.1$ & $41,725$\\
S & $96$ & $423$ & $497$ & $171$ & $13.4$ & $1813.4$\\
\enddata
\end{deluxetable}

\begin{figure*}
    \centering
    \includegraphics[width=0.85\textwidth]{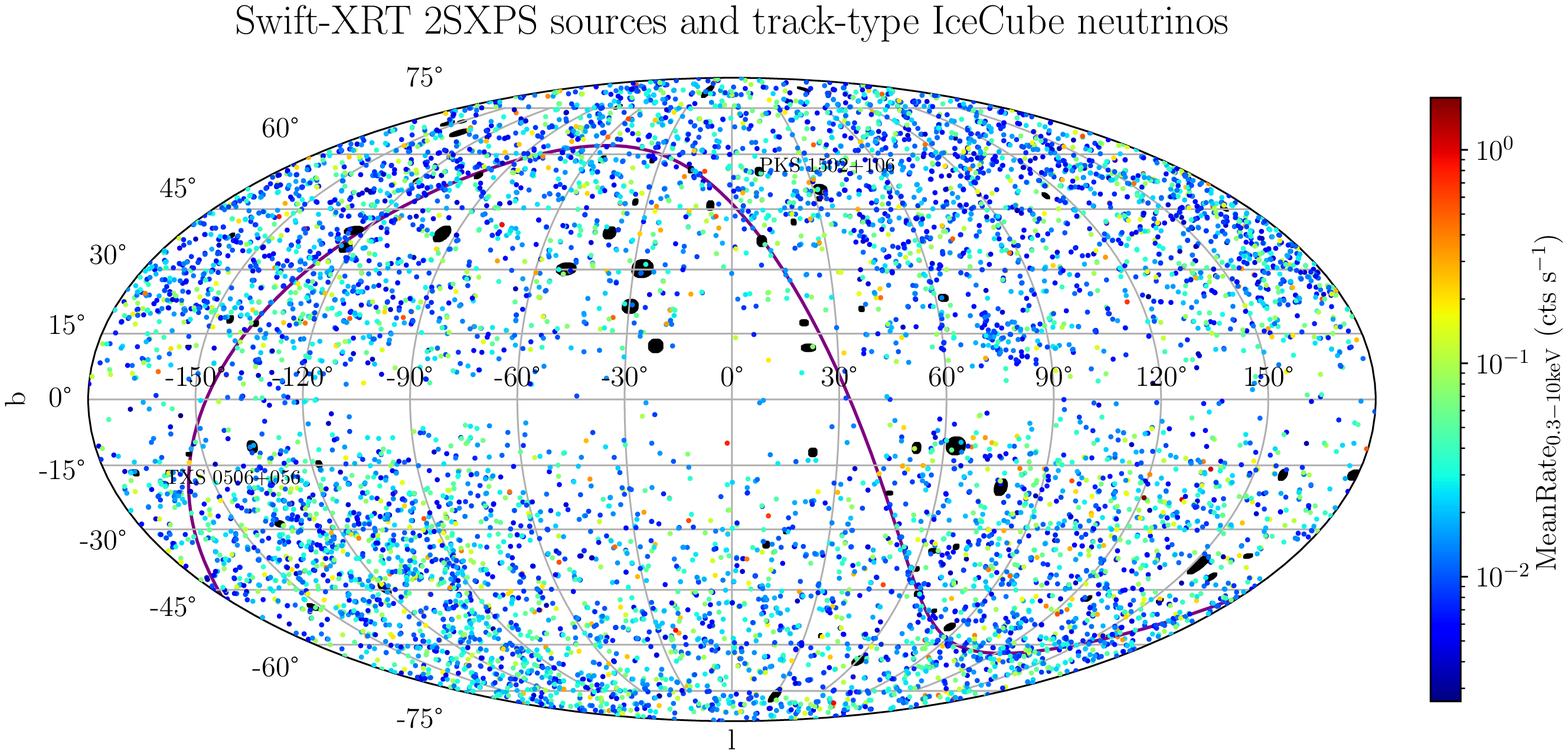}
    \caption{Galactic sky-position ($l$,$b$) and mean rate of the \textit{Swift}-XRT 2SXPS point sources (SNR$\geqq10$, chance of being an AGN is $>0.99$, colored dots) overlaid on the list of the 70 track-type neutrino detection by the IceCube Neutrino Detector (black filled ellipses) compiled by \citet{Giommi2020}. The map is shown in Mollweide projection. The equator is plotted by a purple continuous line. TXS 0506+056 and PKS 1502+106 are marked on the map.}
    \label{fig:xray}
\end{figure*}
\begin{deluxetable*}{ccccccccc}
\tablenum{3}
\tablecaption{Excerpt of the neutrino source candidates from the \textit{Swift}-XRT 2SXPS Catalog. Neutrino ID (1), detection time (2), arrival direction of neutrino (90\% containment, 3-4), 2SXPS source (5), equatorial coordinates (J2000) (6-7), mean rate between 0.3-10 keV (8), signal-to-noise ratio (9). The full table is available in electronic format \href{https://konkoly.hu/staff/kun.emma/table3_swift_fulltable.txt}{here}.\label{table:swiftr}}
\tablewidth{0pt}
\tablehead{
\colhead{$ID_{\nu}$}  & \colhead{$t_{\nu}$} & \colhead{$RA_{\nu}$} & \colhead{$DEC_{\nu}$} & \colhead{$ID_\mathrm{2SXPS-}$} & \colhead{$RA_{Sw}$} & \colhead{$DEC_{Sw}$} & \colhead{$Rm_{Sw}$} & \colhead{$SNR_{Sw}$}\\
 & (MJD) & ($^\circ$) &  ($^\circ$) & & ($^\circ$) &  ($^\circ$)& (cts s$^{-1}$) &  }
 \decimalcolnumbers
\startdata
IC-090813A & 55056.70 & $29.62^{+0.40}_{-0.38}$ & $1.23^{0.18}_{-0.22}$ & J015910.0+010515 & 29.79 & 1.09 &  $0.030\pm0.003$ & 42.0\\
IC-100710A & 55387.54 & $307.53^{+2.70}_{-2.28}$ & $21.00^{2.25}_{-1.56}$ & J203057.1+193612 & 307.74 & 19.60 &  $0.059\pm0.005$ & 93.4\\
IC-100710A & 55387.54 & $307.53^{+2.70}_{-2.28}$ & $21.00^{2.25}_{-1.56}$ & J203031.2+223438 & 307.63 & 22.58 &  $0.020\pm0.004$ & 40.7\\
IC-110521A & 55702.77 & $235.98^{+2.70}_{-1.76}$ & $20.30^{1.00}_{-1.43}$ & J154743.5+205216 & 236.93 & 20.87 &  $0.284\pm0.006$ & 178.6\\
IC-110521A & 55702.77 & $235.98^{+2.70}_{-1.76}$ & $20.30^{1.00}_{-1.43}$ & J154750.7+210352 & 236.96 & 21.06 &  $0.007\pm0.002$ & 13.0\\
\enddata
\end{deluxetable*}

Similar studies were already conducted in the gamma regime, but either with different neutrino lists or with different source catalogs. \citet{Franckowiak2020} studied the IceCube real-time alerts, that selects high-energy
($\gtrsim 100$~TeV) starting and through-going muon track events
\citep[][]{Aartsen2017realtime}, and archival neutrino events that would have passed the same selection criteria. We used the neutrino list by \citet{Giommi2020} that is partially different. When searching for the neutrino source-candidates, \citet{Giommi2020} used the \textit{Fermi}-LAT 3FHL \citep[that reports hard \textit{Fermi}-LAT sources significantly detected in the $10$~GeV--$2$~TeV energy range during the first $7$~years of the \textit{Fermi} mission using the Pass 8 event-level analysis,][]{3HFL2017}, 4LAC \citep[the Fourth Catalog of Active Galactic Nuclei detected by the LAT based on $8$~years of data,][]{Fermi4LAC2020} and 3HSP catalogs \citep[a sample of extreme and high-synchrotron peaked blazars and blazar candidates,][]{Chang2019}, while we employed the whole 10 years catalog of \textit{Fermi}-LAT data.

\subsection{X-ray neutrino source candidates in the \textit{Swift}-XRT 2SXPS catalog}
\label{xray}

The \textit{Swift}-XRT Point Source (2SXPS) catalog covers $3,790$ square degrees on the sky and contains position, fluxes, spectral details and variability information for $206,335$ X-ray point sources. Observations were made with the X-ray telescope on-board the \textit{Neil Gehrels Swift Observatory} between 2005 Jan 01 and 2018 August 01 \citep{Evans2020}. Each source has a detection flag which indicates how likely it is to be a real astrophysical object. We applied several selection criteria: DetFlag=0 (good), SNR(=Count rate/BG rate)$\geqq10$ (in the total 0.3-10 keV energy band). We plot the catalog in the upper panel of Fig. \ref{fig:xray} overlaid on the 90\% containment area of the IceCube track-type neutrino events. Due to the usually wide diversity of sources with different nature in an X-ray catalog such like the 2SXPS (with typically AGN and stars as the most dominant populations), we applied an additional selection criterion that allows only AGN being in our sample. \citet{Tranin2022} applied a robust probabilistic classification of X-ray sources in the \textit{Swift}-XRT and \textit{XMM-Newton} catalogs. We selected such X-ray sources, for which the posterior probability of being an AGN is $>0.99$ according to their probabilistic classification. The resulting sub-sample contains $9,079$ sources having high chance of being an AGN, from which we found $61$ sources to be located within the 90\% containment area of track-type neutrinos, as summarized in Table \ref{table:swiftr}. 

In Table \ref{table:results} we summarize the average, standard deviation, median, minimum and maximum values of the X-ray full band rate in the $0.3$ keV--$10$ keV energy range of the sources in the full 2SXPS catalog, and separately such values for the X-ray $\nu$-sample. The the average flux of the \textit{Swift}-XRT $\nu$-sample emerged as $0.049 \mathrm{cts}s^{-1}$.

\begin{figure*}
    \centering
    \includegraphics[width=0.85\textwidth]{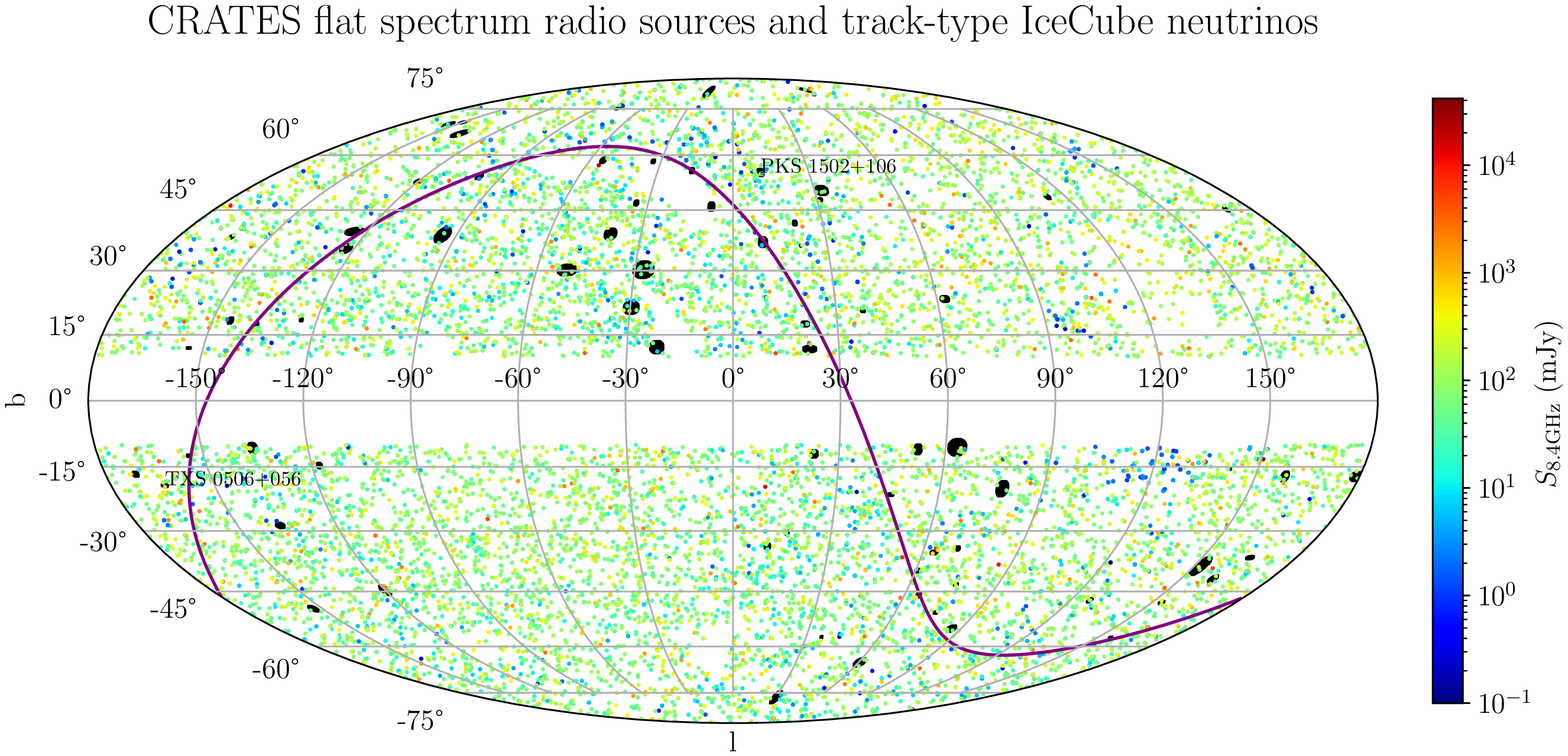}\\
    \caption{Galactic sky-position ($l$,$b$) and 8.4~GHz flux density of the CRATES radio sources (colored dots) and list of the 70 track-type neutrino detection by the IceCube Neutrino Detector (black filled ellipses) compiled by \citet{Giommi2020}. The maps are shown in Mollweide projection. The equator is plotted by a purple continuous line. TXS 0506+056 and PKS 1502+106 are marked on the map.}
    \label{fig:radio}
\end{figure*}
\begin{deluxetable*}{cccccccccc}
\tablenum{4}
\tablecaption{Excerpt of the neutrino source candidates from CRATES catalog of flat spectrum radio sources ($n=100$, out of 14,467 at 8.4~GHz).  Neutrino ID (1), detection time (2), arrival direction of neutrino (90\% containment, 3-4), ID of the CRATES radio source (5), its position (6-7), its detected flux at 4.8 GHz (8) and 8.4 GHz (9) observing frequencies, morphology type (10) as N = No detection at 8.4 GHz, P = point source, S = short jet ($<= 1"$ for VLA maps), L = long jet ($> 1"$). The full table is available in electronic format \href{https://konkoly.hu/staff/kun.emma/table4_crates_fulltable.txt}{here}.\label{table:crates}}
\tablewidth{0pt}
\tablehead{
\colhead{$ID_{\nu}$}  & \colhead{$t_{\nu}$} & \colhead{$RA_{\nu}$} & \colhead{$DEC_{\nu}$} & \colhead{$ID_\mathrm{C}$} & \colhead{$RA_\mathrm{C}$} & \colhead{$DEC_\mathrm{C}$} & \colhead{$S_{4.8}$} & \colhead{$S_{8.4}$} & \colhead{morph}\\
 & (MJD) & ($^\circ$) &  ($^\circ$) & & ($^\circ$) &  ($^\circ$) & (mJy) & (mJy) & 
 }
 \decimalcolnumbers
\startdata
IC-100710A & 55387.54 & $307.53^{2.70}_{-2.28}$ & $21.00^{2.25}_{-1.56}$ & J203332+214613 & 308.38 & 21.77 & 777 & 321.6 & P \\
 &  & & & J203445+205949 & 308.69 & 21.00 & 149 & 88.0 & P \\
 &  & & & J203718+213103 & 309.32 & 21.51 & 74 & 50.0 & P \\
\hline
IC-100925A & 55464.90 & $266.21^{0.58}_{-0.62}$ & $13.40^{0.52}_{-0.45}$ & J174304+132633 & 265.76 & 13.44 & 96 & 51.7 & P \\
\hline
IC-110521A & 55702.77 & $235.98^{2.70}_{-1.76}$ & $20.30^{1.00}_{-1.43}$ & J154534+200654 & 236.39 & 20.11 & 125 & 94.9 & P \\
 &  & & & J154541+204930 & 236.43 & 20.82 & 75 & 73.1 & P \\
 &  & & & J155122+200316 & 237.84 & 20.05 & 198 & - & N \\
\hline
IC-110610A & 55722.43 & $272.49^{1.23}_{-1.19}$ & $35.55^{0.69}_{-0.69}$ & J180812+350104 & 272.05 & 35.02 & 140 & 89.3 & P \\
\hline
IC-110930A & 55834.45 & $267.23^{2.09}_{-1.55}$ & $-4.41^{0.59}_{-0.86}$ & J175138-052032 & 267.91 & -4.66 & 117 & 99.6 & P \\
\enddata
\end{deluxetable*}

\subsection{Radio neutrino source candidates in the CRATES catalog}
\label{radio}

We searched radio sources within the 90\% containment area of neutrinos in the Combined Radio All-Sky Targeted Eight GHz Survey \citep[CRATES,][]{Healey2007}.
The authors have assembled an $8.4$~GHz survey of bright, flat-spectrum ($\alpha > -0.5$, with the convention $S_\nu \sim \nu^{\alpha}$, \citet{Gregorini1984}) radio sources with nearly uniform extragalactic ($|b| > 10$ degrees) coverage for sources brighter than a $4.8$~GHz flux density of $S_\mathrm{4.8~GHz} = 65$ mJy. The catalog is assembled from existing observations, especially the Green Bank 6-cm Radio Source Catalog \citep[GB6, 4.8 GHz][]{Gregory1996}, the Parkes-MIT-NRAO Survey \citep[PMN, 4.8 GHz][]{Wright1994,Griffith1994}, and the Cosmic Lens All-Sky Survey \citep[CLASS, 8.4 GHz][]{Myers2003}. The existing observations were augmented by reprocessing of archival VLA and ATCA data and by new observations to fill in coverage gaps. The resulting catalog provides precise positions, sub-arcsecond structures, and spectral indices for $14,467$ sources at $8.4$~GHz. The number of sources exceeds the number of sources at $4.8$~GHz since in many cases there are multiple $8.4$~GHz counterparts to a single 4.8-GHz source due to the higher frequency and therefore better resolution. There are also $762$ sources for which no $8.4$~GHz counterpart was detected. We plot the catalog in Fig. \ref{fig:radio} overlaid on the 90\% containment area of the IceCube track-type neutrino events. 
We found $87$ ($96$) CRATES sources at $4.8$ GHz ($8.4$~GHz) to be located within the $90$\% containment area of track-type neutrinos, as summarized in Table \ref{table:crates}. In Table \ref{table:results}. we summarize the average, standard deviation, median, minimum and maximum values of the CRATES radio flux densities at $4.8$ GHz and at $8.4$ GHz, and separately such values for the radio $\nu$-samples. The the average flux of the CRATES $\nu$-samples emerged as $544$~mJy at $4.8$ GHz, and $423$~mJy at $8.4$~GHz.

\section{Randomness test of the $\nu$-samples}
\label{sec:randomness}

In this section we investigate whether the selection method based on track-type IceCube neutrinos chooses random $\nu$-samples, or $\nu$-samples with flux properties significantly different from the full catalogs.

The null-hypothesis states a $\nu$-sample represents a sample randomly chosen from the full catalog, while the alternate hypothesis states the average observed flux of a $\nu$-sample is rather special value characteristic to the selection based on the neutrinos.

First we took the angular error of the $70$ neutrinos, randomly generated their RA coordinates (with $0.1$ deci\-mal precision), repeated the source search described in Section \ref{fermisearch}, calculated the average flux contained by the gamma sources, and repeated this process $10,000$ times. We did not scramble the declination of the neutrinos because of the strong variation of the sensitivity of IceCube with declination. Integrating the probability density functions (PDFs) from the observed averages to infinity we get the probability of the corresponding $\nu$-sample to be random under the null hypothesis. We repeated the same process for the \textit{Swift}-XRT 2SXPS catalog, and for the CRATES catalog at $4.8$ GHz and $8.4$ GHz.

\begin{figure}
    \centering
    \includegraphics[angle=270,scale=0.5]{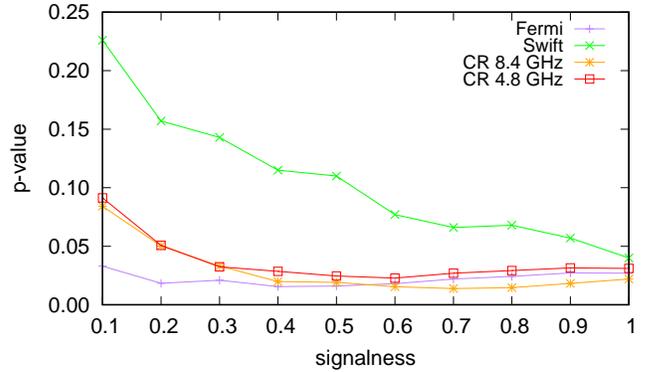}
    \caption{Scan of p-values between the neutrinos and various samples as function of signalness (signalness=1 means every IceCube neutrinos in the sample have astrophysical origin, signalness=0.1 means the chance of the neutrinos to have astrophysical origin is 10\%.)}
    \label{fig:p-values}
\end{figure}

We repeated the tests for different values of the signalness. For example, when we set the signalness to be $0.6$, each of the neutrinos has 60\% probability to be of astrophysical origin relative to the total background rate. Then the chance of a neutrino to enter a cross-match equals the actual signalness of the test. We plot the resulting $p$-values as function of the signalness in Fig. \ref{fig:p-values}. It seems that increasing the signalness of the neutrinos, the $p$-value of the \textit{Fermi-LAT} 4FGL-DR2 and CRATES catalogs first decrease, then the p-values fluctuate about constant values. Increasing the signalness, however, the $p$-value of the \textit{Swift-XRT} 2SXPS continuously decrease without reaching a constant significance level.

We found that there is a reasonable, though not conclusive connection between the neutrinos and the \textit{Fermi}-LAT 4FGL-DR2 gamma sources ($p\approx 0.016$ for $50$\% signalness, with corresponding significance $\sim 2.1\sigma$), as well as between the neutrinos and the CRATES radio sources at $4.8$~GHz and at $8.4$~GHz observational frequencies ($p\approx 0.025$, $p\approx 0.019$, respectively, corresponding significances $\sim 2.0\sigma$, $\sim2.1\sigma$ both for 50\% signalness). We note, the 50\% signalness (s=0.5) represents the cut for which neutrino events detected before 2019 June (since when events with high signalness are unified to Gold and Bronze channels) probably would end up in the Gold channel of IceCube's Real Time Alert System, which is why we chose this value in accordance to previous work \citep{Giommi2020}. A less significant connection emerged between the track-type neutrinos and the \textit{Swift}-XRT 2SXPS catalog ($p\approx 0.11$ for $50$\% signalness, with corresponding significance $\sim 1.2\sigma$). Increasing the signalness of the neutrinos to $100$\%, however, the significance of the subsample derived from the Swift-XRT 2SXPS catalog to be the origin of the neutrinos reaches of a $p$-value of $0.04$ ($\sim 1.8\sigma$).

In the rest of the paper we focus on the radio-neutrino argument. We will present our study focusing on the X-ray and gamma regimes in a subsequent paper.

\section{Investigation of neutrino-radio correlations based on the CRATES catalog}
\label{sec:nuradio}

In this section we estimate how many neutrinos can be explained with the CRATES catalog, assuming the neutrino flux is proportional to the radio flux.

\subsection{Error estimates of the flux density and position of CRATES sources}
\label{subsec:crates_errors}

We needed the flux density uncertainties; however, this information is not directly accessible through the CRATES catalog\footnote{\url{https://vizier.u-strasbg.fr/viz-bin/VizieR-3?-source=J/ApJS/171/61/table5}}. The $4.8$~GHz part of CRATES is composed mainly from the PMN and GB6 surveys. The single-dish observations of the southern sky at $4.85$~GHz are available from the PMN catalog, which covers the region $-87^\circ<\delta<+10^\circ$. Observations were made with the Parkes 64m Radio Telescope. The standard error ($dS$) of the flux densities ($S$) in the PMN catalog can be calculated as \citep[][]{Wright1994}:
 \begin{equation}
    dS=\sqrt{(12.3+0.085\delta)^2+(0.052 S)^2},
\end{equation}
where $\delta$ is the declination. We used this equation to estimate the flux density uncertainties of the radio sources in CRATES south of $\delta<0^\circ$, at $4.8$~GHz. The GB6 catalog of sources covers the declination range $0^\circ<\delta<+75^\circ$, and the observations were made with the 91 m telescope at Green Bank. Since the noises and calibration errors of the map are unknown (to calculate the rms uncertainty in the corrected peak flux density and the source integrated flux density), we fitted a linear function to estimate the error in the flux density of sources in the region $0^\circ<\delta<+75^\circ$ at $4.8$~GHz, where CRATES is based on mostly GB6 survey. The maximum declination of neutrinos $\sim 67.40^\circ$ (IC-140216A) is well south of the maximal declination covered by the GB6 catalog, we do not need to know flux density in the far north region $\delta>75^\circ$.

The CLASS survey contains observations made with the VLA at A configuration at $8.4$~GHz. \citet{Augusto1998} mentioned the typical error on the CLASS flux densities as $5$\%. Since CLASS is a parent catalog of CRATES, and most of the sources in CRATES were observed by the VLA ($11,333$ out of $14,467$), we estimated the errors as $5$\% of the flux densities. 

 In the north region ($0^\circ<\delta<+75^\circ$) the pointing errors of GB6 source positions are estimated as \citep{Gregory1996}: 
\begin{eqnarray}
\epsilon_\alpha\approx 7.5" (\cos \delta)^{1/2},\\
\epsilon_\delta\approx 7.8" (\cos \delta)^{1/2},
\end{eqnarray}
which dominates the position uncertainties for strong sources ($S_p > 50$~mJy). 
In the region of equatorial and far south latitudes ($-87^\circ<\delta<0^\circ$), the position errors of PMN source positions are estimated as \citep{Griffith1993}: 
\begin{eqnarray}
\epsilon_\alpha["]=((1300/\mathrm{S[mJy]})^2+6^2)^{1/2},\\
\epsilon_\delta["]=((1100/\mathrm{S[mJy]})^2+4^2)^{1/2}.
\end{eqnarray}

At $8.4$ GHz, CRATES is largely based on two surveys, the CLASS survey in north, and the (unpublished) PMN-CA survey in the south. The equatorial south was surveyed for gravitational lenses, and X-band data are available in this region. In the CLASS region $0<\delta<+75^\circ, |b|<10^\circ$ the typical angular resolution error is about $0.2"$ \citep{Myers2003}. In the south the PMN-CA survey quotes a typical position error of $0.6"$ in each coordinate for the PMN-CA data.

\subsection{Number counts of CRATES sources in the $z$-samples}

Our first goal was to find the luminosity down to which our sample is complete at a given redshift, which we call "$z$-sample". To construct a complete $z$-sample, we defined the parameter $C(L,z)=N_o(L,z)/N_pL,z)$, where $N_o(L,z)$ ($N_p(L,z)$) is the observed (predicted) number of sources in a given luminosity and redshift bin. 

\begin{figure*}
\includegraphics[scale=0.7]{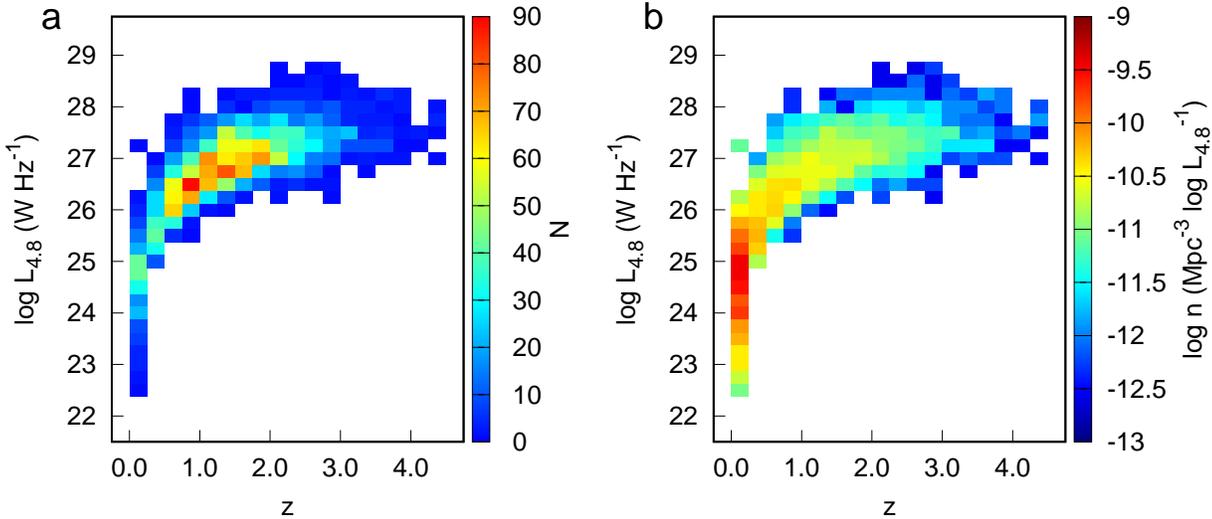}
\caption{(a) Observed number counts in the $z$-sample we assembled from the CRATES catalog at $4.8$~GHz. The coloring corresponds to the number of observed of sources in a given bin, such that blue means low number and red means high number. (b) Number density in the same  $z$-sample. The coloring corresponds to the number density of sources in a given bin, such that blue means low density and red means high density. The binning is $\Delta \log L=0.25$, $\Delta z=0.25$.}
\label{fig:counts}
\end{figure*}

We found the redshift of $3,467$ CRATES sources in the Simbad database using the positional uncertainties summarized in Section \ref{subsec:crates_errors}, and by running Astropy queries \citep{astropy2013,astropy2018}. When more objects were found within the errorbars, the object closest to the CRATES position was assumed (less then $0.5$\% of the cases). The majority of these sources are quasars ($2,262$ sources), BL Lacs ($486$) and blazars ($97$). The maximum of their observed number counts happens between $z\approx 0.75$ and $z\approx 2$ (see Fig. \ref{fig:counts}). It is important to note that nearby blazars like the Markarian sources, galaxies with extraordinary UV brightness compared to normal galaxies, are much closer to us: their redshift ranges from $0.0009$ to $1.912$ with median value $0.0238$, situating them as residents of typically the Local Universe. The First Byurakan Survey (Markarian) Catalog of UV-Excess Galaxies \citep[first paper, and extended list, see][]{Markarian1967,Markarian1989} includes the next classes of objects (with their abundance in brackets): galaxy ($1,251$), starburst galaxy (94), Seyfert type 2 ($65$), galaxy HII region ($26$), LINER ($17$), QSO (13), Seyfert type 1.5 ($12$), BL Lac (3). As it can be seen from this list, only $16$ objects out of $1469$, $13$ QSOs and $3$ BL Lacs are classic radio-loud objects. Therefore, despite their closeness, one does not expect Markarian sources to be a dominant component of neutrino sources observable at radio frequencies, simply because the Markarian galaxies are typically radio-quiet objects \citep[e.g.][]{Biermann1980}.

From the $3,467$ CRATES sources with collected redshifts we kept only flat-spectrum quasars ($2,262$ sources), BL Lacs ($486$) and blazars ($97$), since they follow similar evolution \citep[e.g.][]{Dunlop1990}. These $2,845$ sources were used in the next analysis at the $4.8$~GHz frequency.

The intrinsic luminosity of a source with flux density $S_\nu$ at observing frequency $\nu$, spectral index $\alpha$ (with the convention $S_\nu \sim \nu^\alpha$), seen at luminosity distance $D_L$, redshift $z$, is:
\begin{eqnarray}
    L_\nu=\frac{S_\nu 4\pi D_L^2}{(1+z)^{(1+\alpha)}},
    \label{eq:luminsoity}
\end{eqnarray}
after the $k$-correction $K(z)=(1+z)^{-(1+\alpha)}$.
In Fig. \ref{fig:counts} we show the number counts within our $z$-sample with $\Delta \log L=0.25$, $\Delta z=0.25$ binning. There is luminosity and redshift evolution in the sample, such that the more luminous sources were more abundant in the past. We found the most sources between $\log L\mathrm{(W~Hz^{-1})}\approx26$ and $\log L\mathrm{(W~Hz^{-1})}\approx27.5$, between the redshifts $z\approx0.5$ and $z\approx 2$, such that the peak falls within the bin centered at $\log L_p\mathrm{(W~Hz^{-1})}=26.5$, $z_p\approx0.75$.

The space density of sources as function of the luminosity and redshift can be given in the form \citep[e.g.][]{Ajello2012ApJ}
\begin{equation}
    \frac{d^2N}{dL dz}=\frac{d^2N}{dLd V}\times \frac{dV}{dz}= \phi(L,z) \times \frac{dV}{dz},
\end{equation}
where $\phi (L,z)$ is the radio luminosity function (LF). In this equation $dV/dz$ is the co-moving volume element per unit redshift and unit solid angle \citep[e.g.][]{Porciani2001} which is in a flat universe:
\begin{equation}
    \frac{dV}{dz}=\frac{c}{H_0}\frac{\Delta \Omega_s D_L^2(z)}{(1+z)^2 E(z)},
\end{equation}
where 
\begin{equation}
E(z)=[(\Omega_{g}+\Omega_{\nu})(1+z)^4+\Omega_m(1+z)^3+\Omega_\Lambda]^{1/2},
\end{equation}
$c$ is the speed of the light, $H_0=67.66~\mathrm{km~Mpc^{-1}~s^{-1}}$ is the Hubble parameter at $z=0$, $\Delta \Omega_s$ is the solid angle covered by the survey ($=4\pi$ for the all-sky CRATES), $D_L$ is the luminosity distance, $\Omega_{g}=5.4 \times 10^{-5}$, $\Omega_{\nu}=1.440\times 10^{-3}$, $\Omega_{m}=0.310$, $\Omega_{\Lambda}=0.689$ is the density parameter for the radiation, neutrinos, matter, and dark energy, respectively, taken from \citet{Planck2020}.

First we estimated the LF by employing the $1/V_a$ method of \cite{Avni1980}, which is the generalization of the $1/V_\mathrm{max}$ method of \cite{Schmidt1968} for cases when sufficient number of sources is available in a given survey. If $N$ object appears in the $\Delta \log L \Delta z$ bin element ($L_1< L <L_2$, $z_1<z<z_2$), then the LF is 
\begin{equation}
\phi (L,z)=\frac{1}{\Delta \log L} \sum_{i=1}^N \frac{1}{V_{a}^i},
\end{equation}
where 
\begin{equation}
V_{a}^i= \int _{z_1} ^{\min(z_2,z_\mathrm{max}^i)} \frac{dV}{dz} dz
\end{equation}
is the available volume of the $i$th source, when it is pushed to $z_{max}^i$, where the source would be on the edge of visibility given the lower flux limit of the survey (which is $S_{min}=65$~mJy at $4.8$~GHz for CRATES). The LF $\phi (L,z)$ estimates the space density if all objects in a given population would be 
detectable by the survey with a flux limit. The dimension of $\phi(L,z)$ in our study is $\mathrm{Mpc}^{-3}/ \log L$. The $rms$ error estimate of the LF in each luminosity and redshift bin is calculated by weighting each sources by its contribution \citep[e.g.][]{Marshall1985}:
\begin{equation}
\sigma_\phi(L,z)=\frac{1}{\Delta \log L} \left(  \sum_{i=1}^N \frac{1}{V^2_{a,i}}, \right)^{1/2}.
\end{equation}  
However, if there are ten sources or less in a $\Delta \log L \Delta z$ bin, we used the tabulated upper and lower $84$\% confidence intervals given by \citet{Gehrels1986}, who calculated the Poissonian statistical asymmetrical errors when event rates are calculated from small numbers of observed
events. These intervals correspond to the Gaussian $1\sigma$ errors such that $\sigma_\phi = \phi\times \sigma_N/N$, where $\sigma_N$ is the Poissonian statistical asymmetrical error on the measured number of sources. 
The limiting luminosity distance $D_{L,max}$ of the $i$th source taking into account the CRATES flux limit at $4.8$ GHz can be calculated by keeping $L$ and setting $S_\mathrm{4.8 GHz}=65$ mJy. Then the maximal redshift is calculated by solving 
\begin{equation}
D_\mathrm{L,max}=\frac{c}{H_0} (1+z_\mathrm{max})\int_0^{z_\mathrm{max}} \frac{dz'}{E(z')}
\end{equation}
for $z_{max}$.

\begin{figure}
\includegraphics[scale=0.6,angle=270]{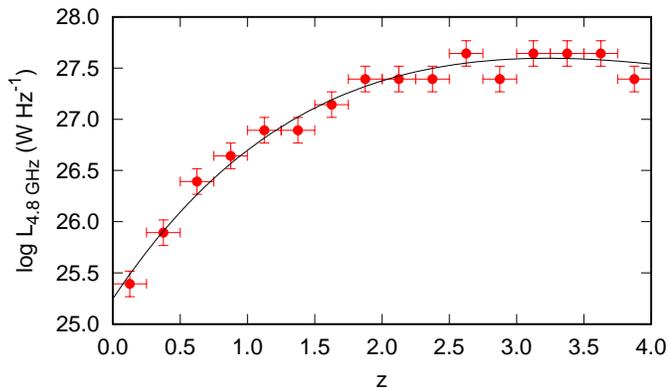}
\caption{The lower limit on the luminosity of sources at 4.8 GHz as function of the redshift. }
\label{fig:comp_l_z}
\end{figure}

Now we estimate the completeness of the $z$-sample derived from the CRATES catalog that comes from our redshift search method. We compared the observed number density to the number density predicted by the $1/V_a$ method, which gives the space density if all sources could have been observed in the survey (with $65$ mJy flux density limit). If the observed number density of sources in the $j$th $\Delta \log L \Delta z$ bin is $n_j$ ($=N_j \Delta \log L^{-1}, \Delta z^{-1}$, the luminosity function is $\phi_j (L,z)$, then
\begin{equation}
{dn_j}{\Delta L\Delta z}=\frac{1}{C_j(L,z)}\phi_j (L,z) dN \frac{dV}{dz} dL dz.
\end{equation}
The completeness parameter $C(L,z)\in [0\div 1]$ takes into account every geometrical and statistical correction factors (e.g. the effective area of the survey) that lead to $N_{j,o} \neq N_{j,p}$, where $N_{j,o}$ ($N_{j,p}$) is the observed (predicted) number of sources in the $j$th bin. In Fig \ref{fig:compl} (see Appendix) we show the resulting completeness parameter $C(L,z)$ in a grid with $\Delta \log L=0.25$, $\Delta z=0.25$.
In Fig. \ref{fig:comp_l_z} we show the luminosity of sources as function of the redshift, over which we capture all sources predicted by the $1/V_a$ method and $C(L,z)=1$. It seems at small redshift the completeness reaches 1 for smaller luminosities, while at larger redshifts we miss the sources with smaller luminosities. We fitted a linear-exponential model in the form of
\begin{eqnarray}
    \log {L^c}_\mathrm{4.8~GHz} (z)=a z \exp{b z}+c
    \label{eq:complete}
\end{eqnarray}
to set the completeness limit in the luminosity redshift plane. The resulting parameters are $a=1.97\pm0.17$, $b=-0.309\pm0.020$,  $c=25.248\pm0.094$ ($\chi^2_\mathrm{red}=0.67$). Sources satisfying $\log L_\mathrm{4.8~GHz} \geqq \log L^c_\mathrm{4.8~GHz}$ ($z$) constitute our complete $z$-sample. The final, complete $z$-sample at $4.8$~GHz contains $1,423$ sources.

\begin{figure*}
    \centering
    \includegraphics[width=0.45\textwidth]{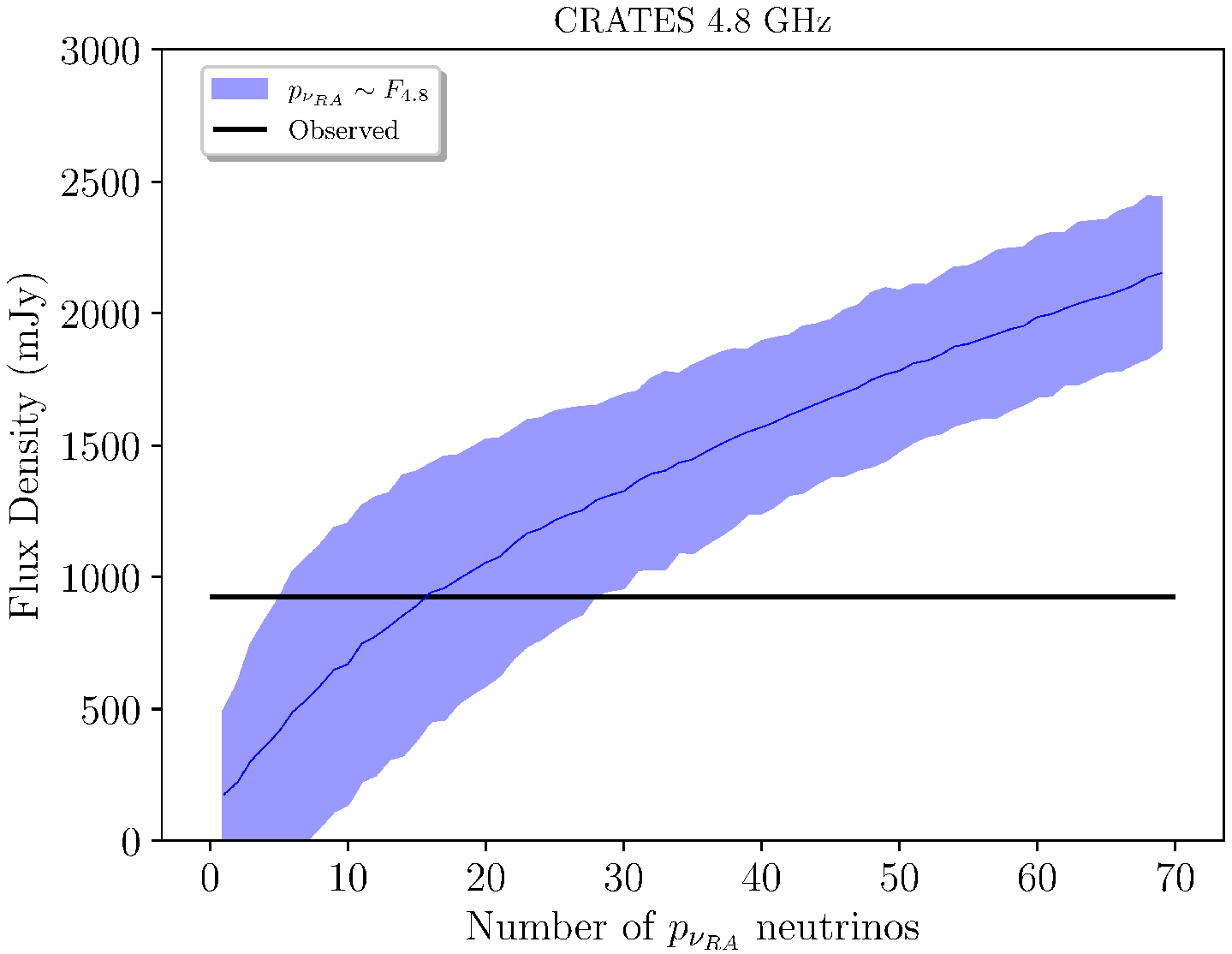}
    \includegraphics[width=0.45\textwidth]{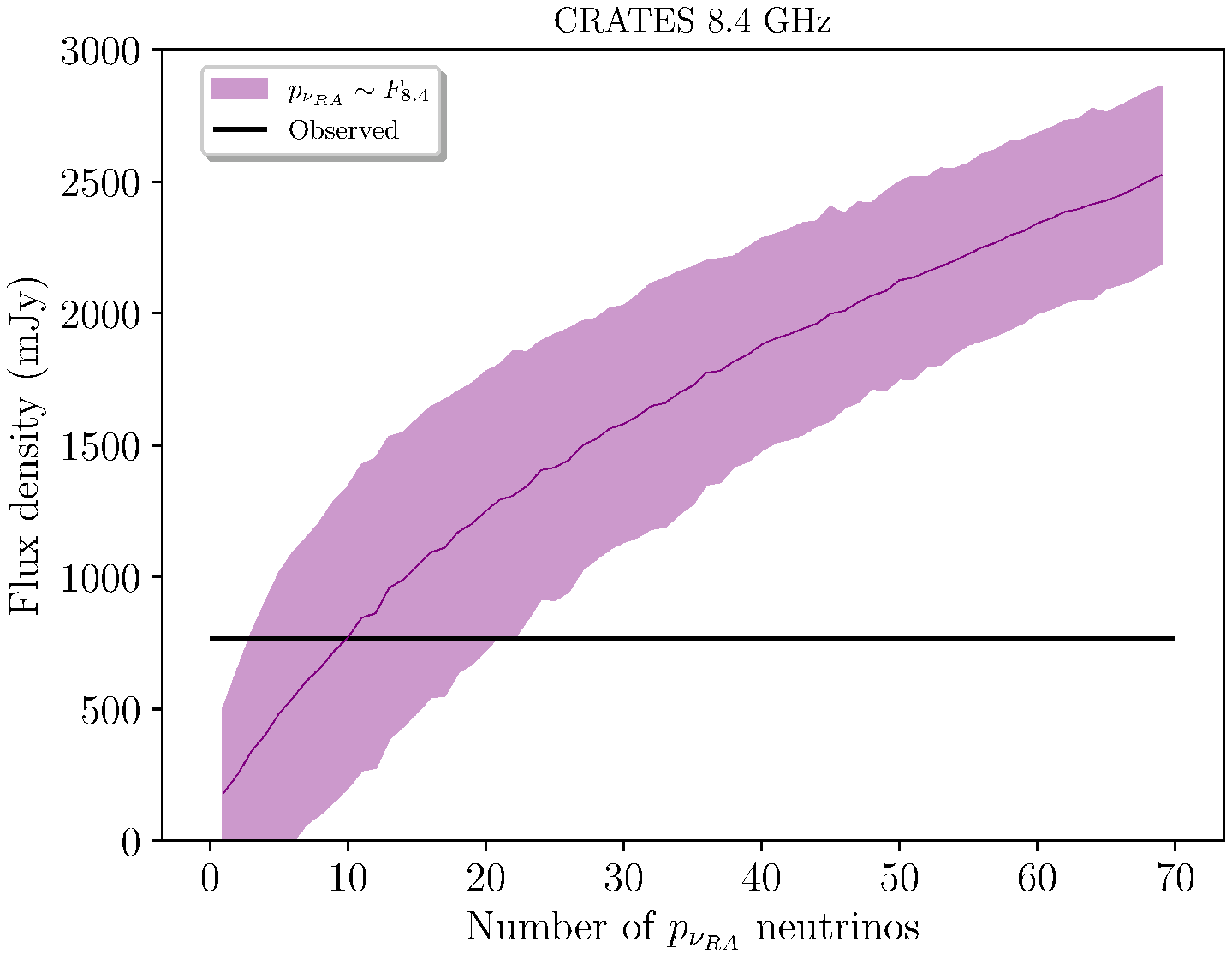}\\
     \caption{Sample average and 90\% confidence interval of $4.8$~GHz (left) and $8.4$~GHz (right) CRATES sources contained by $70$ neutrinos from $1000$ trials (100\% signalness of the neutrinos), such that the probability of the assigned direction of $N$ neutrinos depend on the radio flux, and $70-N$ neutrinos have scrambled right-ascension and the original declination (since the strong declination-sensitivity of the IceCube Detector).}
    \label{fig:confplots}
\end{figure*}
Next we transformed the $8.4$ GHz flux densities to $4.8$ flux densities, such as
\begin{equation}
    S_{\nu_1}=S_{\nu_2} \left(\frac{\nu_1}{\nu_2}\right)^{\alpha},
\end{equation}
then calculated the luminosities by using Eq. \ref{eq:luminsoity} and constrained the complete sample at $8.4$ GHz based on Eq. \ref{eq:complete}. The complete $z$-sample at 8.4 GHz contains $1,428$ sources. We also calculated the $8.4$~GHz luminosities.

We repeated the catalog-matching taking into account only the complete $4.8$ GHz and $8.4$~GHz $z$-samples derived from the CRATES catalog, and the $70$ IceCube track neutrinos from \citet{Giommi2020}. We found that $13$ neutrinos contain $14$ CRATES sources at $4.8$~GHz, and $14$ neutrinos contain $15$ CRATES sources from the $z$-sample. The average flux density of the neutrino-bright CRATES sources is $925$~mJy ($768$~mJy) at $4.8$~GHz ($8.4$~GHz) when we assume 100\% signalness, and $956$~mJy ($762$~mJy) at $4.8$ GHz ($8.4$~GHz) when we assume 50\% signalness of the neutrinos. The 50\% signalness was taken into account as described in Section \ref{sec:randomness}.

\subsection{Does the probability to detect a neutrino relate to the radio flux?}
\label{discussion}

After constraining a sub-sample of the CRATES catalog complete in radio luminosities, we estimated that how many neutrinos could be explained with this sub-sample at $4.8$~GHz and $8.4$~GHz, if the probability to detect a neutrino is assumed to be proportional to the radio flux at these observational frequencies. With this we assume that neutrinos are emitted whenever a source is in an active stage, and assume that this probability scales with flux at radio frequencies. We also assume once a source is in an active stage, the probability to actually detect it runs with the $k$-corrected (radio) flux $F=L/(4 \pi {D_L}^2)$. 

We assigned the sky coordinates of $N$ CRATES sources from the 4.8 GHz complete z-sample to $N$ neutrinos keeping the (randomly assigned) angular error of the neutrinos. The probability of the assigned direction depends on the radio flux of the given radio source. The right-ascension of the remaining $70-N$ neutrinos was scrambled (with $0.1$ deci\-mal precision). We did the source finding as described in Section \ref{sec:matches}, and calculated the average flux density contained within the containment area of $70$ neutrinos. We repeated the same process $10,000$ times, and then calculated the mean of the $10,000$ sample average flux densities and its $90$\% confidence interval. We then run $N$ from $N_\mathrm{min}=1$ to $N_\mathrm{max}=70$. We then repeated the whole process employing the $8.4$~GHz complete $z$-sample.

We plot the mean of the $4.8$~GHz and $8.4$~GHz sample average flux densities and their 90\% confidence intervals in Figure \ref{fig:confplots}, where we run $N$ from $1$ to $70$. We see that higher the number of neutrinos with flux density-driven directions higher the sample average, as more and more neutrino directions equal the direction of the brightest CRATES sources and the catalog matching indeed finds these CRATES sources.

Taking into account the observed average flux densities ($925$~mJy at $4.8$~GHz and $768$~mJy at $8.4$~GHz), we find that the CRATES catalog can explain $15.7^{+12.6}_{-10.5}$ neutrinos at 4.8 GHz and $10.0^{+11.3}_{-7}$ neutrinos at $8.4$~GHz (90\%~C.L.), when the probability to detect a neutrino is assumed to be proportional to the radio flux. We repeated the test by assuming 50\% astrophysical signalness of the neutrinos. In that case we found the CRATES catalog can explain $6.6^{+11.9}_{-5.25}$ neutrinos at 4.8 GHz and $4.4^{+10.3}_{-3.5}$ neutrinos at $8.4$~GHz (90\%~C.L.).

\section{Summary of results}
\label{sec:results}

\subsection{Gamma-ray, X-ray and radio position matches with IceCube neutrino directions and probability of the $\nu$-samples}

In Section \ref{sec:matches}. we found $29$ \textit{Fermi}-LAT 4FGL DR-2 gamma-ray sources (out of $5,787$), $61$ \textit{Swift}-XRT X-ray point sources (out of $9,097$, with $SNR\geqq10$ and considering only AGN-type objects), and $87$ ($96$) CRATES radio sources at $4.8$~GHz ($8.4$~GHz) located within the 90\% containment area of the $70$ IceCube track-type neutrino events listed by \citet{Giommi2020}. 

We found that out of the 70 IceCube track-type neutrino events, $53$ cover at least one gamma, or one X-ray, or one radio source. Out of these $53$ neutrinos, $15$ contain at least one Fermi, one \textit{Swift}-XRT and one CRATES source. One neutrino contains at least one \textit{Fermi} and one \textit{Swift} source. Five neutrinos contain at least one \textit{Fermi} and one CRATES sources, and also five neutrinos contain at least one \textit{Swift} and one CRATES sources. One neutrino contains only \textit{Fermi}-LAT, 7 contain only \textit{Swift}, and 18 neutrinos contain only CRATES sources. 

With hypothesis tests we concluded the $70$ IceCube track-type neutrinos select a special sample from the \textit{Swift}-XRT 2SXPS catalog at a $1.2\sigma$ significance level, when we assume 50\% signalness of the neutrinos. We found that there is a reasonable connection between the neutrinos and the \textit{Fermi-LAT} 4FGL-DR2 gamma-ray sources ($2.1\sigma$), as well as between the neutrinos and the CRATES radio sources at $4.8$~GHz ($2\sigma$), and at $8.4$~GHz ($2.1\sigma$) assuming 50\% signalness.

\subsection{Test of radio flux dependence of neutrino directions based on complete $z$-samples}

After constraining $z$-samples complete in $4.8$~GHz and $8.4$~GHz radio luminosities, we found that $13$ neutrinos, out of the $70$ IceCube track neutrinos from \citet{Giommi2020}, contain $14$ CRATES sources at $4.8$~GHz, and $14$ neutrinos contain $15$ CRATES sources from the $z$-sample. Their observed average flux density is $925$~mJy ($768$~mJy) at $4.8$ GHz ($8.4$~GHz). We concluded that the complete $z$-samples derived from the CRATES catalog can explain $15.7^{+12.6}_{-10.5}$ neutrinos at $4.8$ GHz and $10.0^{+11.3}_{-7}$ neutrinos at $8.4$~GHz (90\% C.L.), when the probability to detect a neutrino is assumed to be proportional to the radio flux. We found the complete $z$-samples derived from the CRATES catalog can explain $6.6^{+11.9}_{-5.25}$ neutrinos at 4.8 GHz (between $4$\% and $53$\% of the neutrinos) and $4.4^{+10.3}_{-3.5}$ (between $3$\% and $42$\% of the neutrinos) at $8.4$~GHz (90\%~C.L.), when assuming 50\% astrophysical signalness of the high-energy neutrinos.

\section{Discussion}
\label{sec:discussion}

\citet{Aartsen2017} presented a likelihood analysis searching for cumulative neutrino emission from blazars in
the 2nd \textit{Fermi}-LAT AGN catalog (2LAC) using the 59-string
(IC-59), 79-string (IC-79), and 86-string (IC-86) configurations of the IceCube detector recorded between May 2009 and April 2012 (only track-like events). They constrained the maximum contribution of 2LAC blazars to the neutrino flux observed in the above time period to 27\% or less.

\citet{Pinat2017} presented their search for extended sources of neutrino emission with 7 years of IceCube data (IC40, IC59, IC70, IC86-I, IC86-II, IC86-III, IC86-IV). They found the five maps with simulated source extension of $1\degr$, $2\degr$, $3\degr$, $4\degr$, $5\degr$, all are consistent with background only hypothesis.

\citet{Hooper2019} studied one year of track data (IC86-2011) applying spatial match between this data set, the 3LAC and the Roma BZCAT catalogs. They also searched spatial and temporal match between the above neutrino data set and the \textit{Fermi} Collaboration's All-Sky Variability Analysis. They found no evidence that blazars generate a significant flux of high-energy neutrinos, and found that no more than 5\%-15\% of the diffuse neutrino flux (IC86-2011) can originate from this class of objects.

\citet{Smith2021} used three years of IceCube
data (IC79, IC86-2011, IC86-2012, 2010-2012) to search for evidence of neutrino emission from the AGN in the fourth catalog of AGNs detected by the \textit{Fermi}-LAT (4LAC). They found at the 95\% confidence level, that blazars can produce no more than 15\% of IceCube’s observed flux (2010-2012), and that for non-blazar AGN it remains possible that this class of sources could produce the entirety of the diffuse neutrino flux observed by IceCube in three years.

We employed the full 4FGL-DR2 catalog (ten years of \textit{Fermi} point-like sources) to search the sources of the 70 neutrinos. We found a $\sim 2\sigma$ connection between this catalog and the $70$ track-type neutrino events. It has a similar meaning to the above findings: blazars, dominant population of the 4FGL catalogs, contribute to the IceCube neutrino flux but either the contribution is subdominant, or possible in-source gamma attenuation of neutrino sources complicates the picture.

What we can learn from these results is that there are indications for a correlation between the high-energy neutrinos measured with IceCube and the radio, X-ray and GeV gamma-ray catalogs under investigation in this paper. The question now is how to proceed in order to strengthen these potential correlations to move toward a $5\sigma$ confirmation. Knowledge on the sources and from other multimessenger channels needs to be exploited: what we need to understand is how the different messengers and energy channels are correlated. As an example, when looking at the broader picture of gamma-ray vs. neutrino emission, we are in this paper testing the correlation between GeV gamma-rays and neutrinos. However, it is clear from different observation channels \citep[e.g.][]{Kun2021} that the GeV gamma-ray intensity seems to be significantly lower than what is predicted in the typical correlated intensities of high-energy gamma-rays and neutrinos that arise from the decay of pions. This problem prevails both when looking at the diffuse signal \citep{Ahlers2015}, but also when looking at potential sources like NGC~1068 with a $3\sigma$ evidence for neutrino emission in the $10$ year point source sample \citep[][]{Aartsen2020}. One solution to this problem is that the secondaries are produced in an environment of high photon or gas densities. In the case of NGC~1068, it is likely that the emission happens near the Corona of the accretion disk of this active galaxy \citep[see e.g.][]{Eichmann2021}. As for blazars, a time-dependent modeling is certainly necessary in order to understand the complex multimessenger lightcurves, where the evidence of two very different flaring behaviors of TXS0506+056 is a first indication that no flare is like another \citep[][]{ICTXS2018a,ICTXS2018b}. Different AGN propagation codes have been developed in recent years in order to tackle this problem of time dependence, like ATHE$\nu$A \citep{Dimitrakoudis2012}, Böttcher \citep{Boettcher2013}, Paris \citep{Cerruti2015}, AM$^3$ \citep{Gao2017}, CRPropa \citep{Hoerbe2020}. Most importantly, progress in the understanding of the astrophysical properties of AGN is necessary to really be able to produce quantitative models. Knowing things like the gas and photon field distributions, magnetic field structure and strength from astrophysical observations help to tighten the parameter space and thus make the results of the modeling more reliable. Here, a crucial role lies in the understanding of the plasma, and acceleration and transport process of cosmic rays \citep[][]{Sironi2011,BeckerTjus2022}.

\section{Conclusions}
\label{sec:conclusion}

We examined if there is a correlation between the list of the $70$ IceCube track neutrinos compiled by \citet{Giommi2020}, the gamma-ray \textit{Fermi}-LAT 4FGL-DR2, the X-ray \textit{Swift}-XRT 2SXPS, and the radio CRATES catalogs. By collecting redshifts of sources in the CRATES catalog and deriving sub-samples complete in luminosities, we investigated whether the neutrino flux depends on the radio flux of the complete subsample. In the followings we list the most important takeaways from our recent work.
\begin{itemize}
    \item 
    We found similar levels of correlation between the $70$ IceCube neutrinos and gamma-ray, X-ray and radio source catalogs. Each of these catalogs correlated with neutrino sources at the $1.2$--$2.1 \sigma$.
    \item The sub-sample of CRATES radio sources complete in radio luminosity can explain between $4$\% and $53$\% of the neutrinos at 4.8 GHz and between $3$\% and $42$\% of the neutrinos at $8.4$~GHz ($90$\%~C.L. and 50\% signalness), when the probability to detect a neutrino is assumed to be proportional to the ($k$-corrected) radio flux. This result is consistent with the contribution of AGNs to IceCube's neutrinos based on individually identified sources \citep{Bartos2021}.
\end{itemize}

The next step is to disentangle the correlation between the three photon energies to understand which one is most relevant to neutrinos.
 
\begin{acknowledgments}
 E.K. thanks the Hungarian Academy of Sciences for its Premium Postdoctoral Scholarship. JT and AF acknowledge support from the German Science Foundation DFG, via the Collaborative Research Center \textit{SFB1491: Cosmic Interacting Matters - from Source to Signal}. I.B. acknowledges the support of the Alfred P. Sloan Foundation and NSF grants PHY-1911796 and PHY-2110060.
\end{acknowledgments}

\begin{figure*}
\includegraphics[scale=0.75]{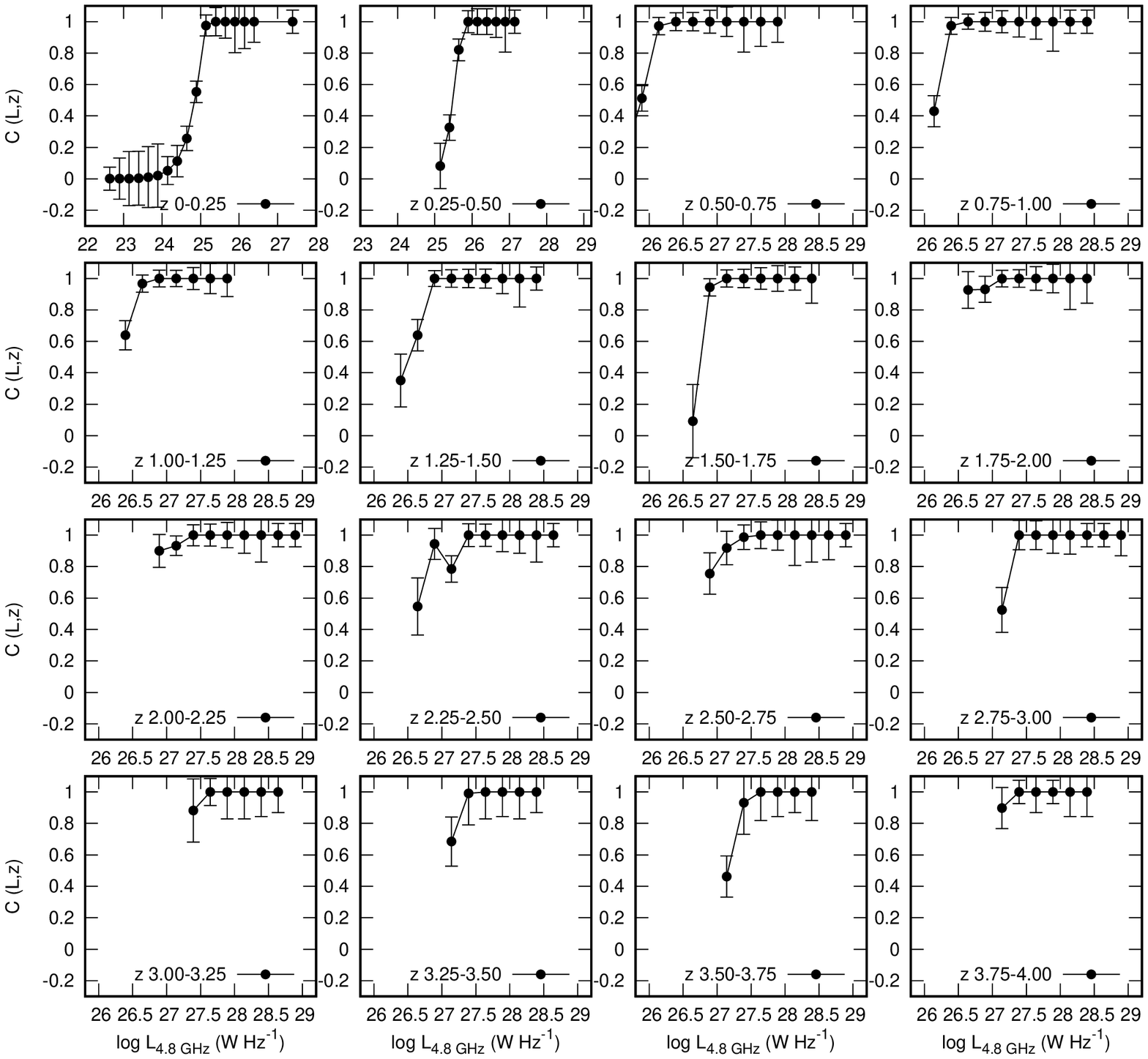}
\caption{The black dots with errorbars show the completeness parameter C (L,z) in our z-sample as function of the logarithm of the 4.8 GHz luminosity for different redshift bins ($2845$ sources with redshifts: $2262$ quasars, $486$ BL Lacs, $97$ blazars).}
\label{fig:compl}
\end{figure*}

\end{document}